\documentclass[onecolumn,manuscript,showpacs,amsmath,amssymb]{revtex4}
\usepackage{}
\usepackage{txfonts}
\usepackage{pifont}
\usepackage{amssymb}
\usepackage{dcolumn}
\usepackage{amsmath}
\usepackage[dvips]{epsfig}

\makeatletter
\def\btt#1{\texttt{\@backslashchar#1}}%
\DeclareRobustCommand\bblash{\btt{\@backslashchar}}%
\makeatother

\begin{document}
\title{Orbital density wave induced by electron-lattice coupling in orthorhombic iron pnictides}
\author{Da-Yong Liu$^{1}$, Ya-Min Quan$^{1}$, Dong-Meng Chen$^{2}$, Liang-Jian Zou$^{1,
        \footnote{Correspondence author, Electronic mail:
           zou@theory.issp.ac.cn}}$, and Hai-Qing Lin$^{3}$}
\affiliation{ \it $^1$ Key Laboratory of Materials Physics, Institute of Solid State Physics,
Chinese Academy of Sciences, P. O. Box 1129, Hefei 230031, People's Republic of China\\
\it $^2$ College of Physics Science and Technology, China University of Petroleum,
DongYing 257061, People's Republic of China\\
\it $^3$ Department of Physics, Chinese University of Hong Kong, Shatin, New Territory, Hong Kong, China\\}
\date{Sep 20, 2010}

\begin{abstract}
In this paper we explore the magnetic and orbital properties closely
related to a tetragonal-orthorhombic structural phase transition in
iron pnictides based on both two- and five-orbital Hubbard models.
The electron-lattice coupling, which interplays with electronic
interaction, is self-consistently treated. Our results reveal that
the orbital polarization stabilizes the spin density wave (SDW)
order in both tetragonal and orthorhombic phases. However, the
ferro-orbital density wave (F-ODW) only occurs in the orthorhombic
phase rather than in the tetragonal one. Magnetic moments of Fe are
small in the intermediate Coulomb interaction region for the striped
antiferromangnetic phase in the realistic five orbital model. The
anisotropic Fermi surface in the SDW/ODW orthorhombic phase is well
in agreement with the recent angle-resolved photoemission
spectroscopy experiments.
These results suggest a scenario that the magnetic phase transition
is driven by the ODW order mainly arising from the electron-lattice
coupling.

\end{abstract}

\maketitle

\section{INTRODUCTION}

The recent discovery of superconductivity in RFeAs$_{1-x}$O$_{x}$
(R=La, Ce, Sm, Pr, Nd, etc.) with high transition temperature has
attracted extensive interest \cite{JACS130-3296}. The parent
compound LaFeAsO shows strong anomalies near 150 K in resistivity,
magnetic susceptibility, specific heat, {\it etc.}. It is suggested
that the ground state is a spin-density-wave (SDW) ordered state
with a striped antiferromagnetic (S-AFM) configuration
\cite{EPL83-27006}. The predicted magnetic structure was then
confirmed by a subsequent neutron diffraction experiment
\cite{Nature453-899}. Interestingly, the neutron diffraction data
indicated that a subtle structural distortion occurs first at
$T_{s}$ $\thicksim$ 155 K, and the SDW long-range order establishes
at a slightly lower temperature, $T_{SDW}$ $\thicksim$ 137 K
\cite{Nature453-899}. This phenomenon also has been generically
found in other 1111 systems of the iron pnictides, $e.g.$, NdFeAsO
($T_{s}$ and $T_{SDW}$ $\thicksim$ 150 K and 141K, respectively.)
\cite{PRB78-064515,PRL101-257002}, and in the 111 systems, $e.g.$
NaFeAs with 52 K and 41 K \cite{PRL102-227004}, and in the 11
systems, $e.g.$ FeTe with 87 K and 75 K \cite{PRL102-247001}.
However in 122 systems, the structural and magnetic phase
transitions spontaneously happen, $i.e.$, at 142 K for
BaFe$_{2}$As$_{2}$ \cite{PRL101-257003}, and at 205 K for
SrFe$_{2}$As$_{2}$ \cite{PRB78-100504,PRB78-212502,PRL101-167203},
{\it etc.}. Although the structural phase transition occurs before
the magnetic ordering is formed, it is difficult to distinguish
either the instability of the electronic structure or the
orthorhombic lattice distortion plays an essential role.
Especially, unless there exists strong magneto-elastic coupling, it
is very strange that the structural transition and the magnetic
transition are so closely related.

The SDW order is suppressed by electron or hole doping, and the
superconductivity is triggered in doped iron pnictides
\cite{EPL83-27006}. The closeness of the superconducting phase to
the SDW instability implies that the AFM fluctuations play a key
role in the superconducting pairing mechanism. Therefore,
investigating the origin of the SDW instability in the parent
compound is an essential step to understand the microcosmic origin
of the superconductivity.
Up to date, two main mechanisms were proposed for addressing the
SAFM ordering in iron pnictides. The first one suggested that the
SAFM results from the Fermi-surface nesting of itinerant electrons
\cite{EPL83-27006, PRL101-057003, PRB77-220506R, PRL100-237003,
PRL101-087004}, which is itinerant SDW character with modulation
wavevector Q=($\pi$,0). Alternatively, it was proposed that the
superexchange interaction mediated through the off-plane As atom
dominates the spin configuration formation \cite{PRL101-057010,
PRL101-076401, PRB78-224517, PRB77-224509, PRB78-020501R,
PRL101-126401}. In such a local picture, the SAFM ordering arises
from the competition between the nearest-neighbor (NN) and next
nearest-neighbor (NNN) spin couplings, when the NNN exchange
coupling becomes larger than half of the NN exchange interaction.
Whether an itinerant picture \cite{EPL82-37007, JPCM20-425203,
PRB77-220503} or a localized superexchange mechanism
\cite{PRL101-076401, PRB78-020501R, PRB77-224509, PRL101-057010}
responsible for the SDW ordering is still a hot debate for
Fe-based layered systems \cite{PRB78-085104}.

On the other hand, it is now believed that iron pnictides are the
multi-orbital systems, with Fe$^{2+}$ ion in a tetrahedral crystal
field (CF). The energy level splittings of the five 3d orbitals are
small, in a magnitude order of 0.1 eV \cite{PRB77-220506R,
PRL101-076401}. Thus the splitting is sensitive to the lattice
distortion.
It is found that in the high-temperature tetragonal phase, the xz
and yz orbitals are degenerate \cite{PRL100-226402}. While we expect
such a degeneracy is
removed in the low-temperature orthorhombic phase.
Up to date, the minimal two-orbital model with half-filling
\cite{PRB77-220503,PRB79-104510,JPSJ78-083704}, three-orbital model
with a filling of one third (i.e., two electrons in three orbitals)
\cite{PRB78-144517} and a filling of two thirds (i.e., four
electrons in three orbitals)
\cite{PRB81-014511,PRL105-096401,PRB81-180514R,PRB82-104508},
four-orbital model with half-filling \cite{PRB79-104510}, and full
five-orbital model (with six electrons in five orbitals) in the two
dimensional (2D) case
\cite{NJP11-025016,PRL101-087004,PRB80-094531,PRB81-144509,
PRB81-180514R,PRB82-104508,PRL104-227201,arXiv1007.1949} and
three dimensional (3D) one \cite{PRB80-104503,PRB81-214503}, were
proposed for addressing the low-T electronic, magnetic and optical
properties of the iron pnictides.
However, in most of previous works \cite{PRB79-104510,JPSJ78-083704,
PRB81-014511,PRL105-096401,PRB82-104508,PRB81-180514R,
PRL104-227201,arXiv1007.1949}, the orthorhombic CF splitting was
neglected. Only a few authors suggested the role of the orthorhombic
CF splitting \cite{PRB80-224504,PRB80-224506,PRB82-100504R}, but the
magnitude of the CF splitting was not self-consistently determined.
Therefore, whether the degeneracy of the orbitals and the
orthorhombic distortion play an essential role on the groundstate
properties or not is still not very clear.

The remove of the degeneracy of the xz/yz orbitals may be associated
with the orbital ordering (OO) in iron pnictides.
%
%
In a recent angle-resolved photoemission spectroscopy (ARPES)
experiment on BaFe$_{2}$As$_{2}$, it was found that
magnetostructural transition is accompanied with orbital-dependent
modifications in the electronic structure \cite{PRL104-057002},
which is an evidence of the orbital polarization or OO. Especially,
the infrared phonon anomaly in BaFe$_{2}$As$_{2}$ was also thought
to be the consequence of the ordering of the orbital occupation
\cite{PRB80-180502R}.
In addition, the magnetic \cite{nphys5-555}, resistance
\cite{PRB81-214502,science329-824, arXiv1002.3801} and optical
conductivity \cite{arXiv1007.2543} anisotropies were observed in the
experiments, which were attributed to be the evidence of the
orbital polarization.
%
Meanwhile, the possible influence of the orbital polarization has
also been investigated by many other experimental and theoretical
works, including local-density approximation (LDA)
\cite{PRL103-267001,PRL104-227201,PRB81-180514R}.
The OO concept based on the strong correlation and localized
picture, which is usually applied for classical insulator materials,
had been proposed to being responsible for the SDW order
\cite{PRB79-054504}. However, the iron pnictides, being a bad metal and in
the moderate correlation regime
\cite{nphys5-647}, it doesn't seem to fall
into this class material.
Therefore, the orbital density wave (ODW) based on itinerant
scenario can be applied on these moderate correlation systems
\cite{CPL26-097501,PRB80-024512,EPL88-17004,arXiv1003.1660}.

Although some theoretical works proposed
the OO scenario \cite{PRL103-267001,PRL104-227201}, to interpret
the origin of the orbital polarization of iron pnictides observed
in experiment,
other authors suggested that these results are mainly driven by the
magnetic ordering in absence of the OO \cite{PRB81-180514R}.
Recently, the strong electron-lattice (e-l) interaction associated
with the orthorhombic distortion in BaFe$_{2}$As$_{2}$ has been
suggested to take responsiblility for the structural phase transition
\cite{arXiv1008.1479}, which contradicts with the scenario that
the OO induced by purely electron-electron (e-e) interaction drives
the structural phase transition \cite{PRL103-267001}.
Therefore, the origin of the orbital polarization and its consequence
are still an open and debating question. As an alternative, we
suggest recently that the ODW favors the existence of the SDW in
iron pnictides \cite{CPL26-097501}. In this paper, we propose that
the SDW ordering is driven by the density wave-type OO state in iron
pnictides. We find that such an OO phase is stabilized only in the
presence of the orthorhombic distortion. The OO physics in five
orbital model is similar to the degenerate two-orbital picture due
to the Jahn-Teller-type distortion in the itinerant background.
This paper is organized as follows: a model Hamiltonian and
mean-field approximation are described in {\it Sec. II}; then the
numerical results and discussions are presented in {\it Sec. III};
the last section is devoted to the remarks and summary.

\section{Effective Hamiltonian and Ordering Parameters}

We start with an extend multi-orbital Hubbard model Hamiltonian including
both the e-e and e-l interactions,
\begin{eqnarray}
  H=H_{0}+H_{I}+H_{e-l}.
\end{eqnarray}
Here H$_{0}$ describes the kinetic energy term,
\begin{eqnarray}
   H_{0} &=&
   \sum_{\substack{i,j\\ \alpha,\beta,\sigma}}t_{ij}^{\alpha\beta}
   C_{i\alpha\sigma}^{+}C_{j\beta\sigma}-\mu\sum_{i\alpha\sigma}n_{i\alpha\sigma}
\end{eqnarray}
where $C_{i\alpha\sigma}^{+}$ creates an electron on site i with
orbital $\alpha$ and spin $\sigma$, $t_{ij}^{\alpha\beta}$ is the
hopping integral between the i site with $\alpha$ orbital and the j
site with $\beta$ orbital, and $\mu$ is the chemical potential
determined by the electron filling. The electronic interaction part
reads,
\begin{eqnarray}
  H_{I} &=& U\sum_{\substack{i, \alpha}}n_{i\alpha\uparrow}n_{i\alpha\downarrow}
  +U^{'}\sum_{\substack{i\\ \alpha\ne\beta}}n_{i\alpha\uparrow}n_{i\beta\downarrow}
  +(U^{'}-J_{H})\sum_{\substack{i,\sigma}}n_{i1\sigma}n_{i2\sigma}
\nonumber\\
  &&-J_{H}\sum_{i}(C_{i1\uparrow}^{+}C_{i1\downarrow}C_{i2\downarrow}^{+}C_{i2\uparrow}
  +h.c.)+J_{H}\sum_{i}(C_{i1\uparrow}^{+}C_{i1\downarrow}^{+}C_{i2\downarrow}C_{i2\uparrow}
  +h.c.)
\end{eqnarray}
where U(U$^{'}$) denotes the intra-(inter-)orbital Coulomb repulsion
interaction and J$_{H}$ the Hund's coupling. We take
U$^{'}$=U-2J$_{H}$ throughout this paper.

The e-l interaction is depicted as
\begin{eqnarray}
   H_{e-l} &=& \frac{\delta}{2}\sum_{\substack{i\sigma}}
   (n_{i1\sigma}-n_{i2\sigma})e^{i\overrightarrow{Q}_{o}\cdot\overrightarrow{R}_{i}}
   +\frac{\lambda}{2}\sum_{\substack{i}}\delta^{2}_{i}
\end{eqnarray}
where $\delta$ is the CF level splitting of the xz and yz orbitals
induced by orthorhombic distortion, and $\overrightarrow{Q}_{o}$ is
the ODW vector, $\lambda$ is a constant coefficient. Since the
metallic iron pnictides do not fall into conventional OO scenario
for insulators, it is only a SDW-type spin ordering and ODW-type OO
for complicated metals \cite{CPL26-097501}. In the presence of the
SDW and ODW order, we adopt the following mean-field approximation
in real space to decouple the particle-particle interaction terms in
Eq. (3),
\begin{eqnarray}
   <n_{i\alpha\sigma}> &=& \frac{1}{4}[n_{12}+\sigma m_{s}e^{i\overrightarrow{
   Q}_{s}\cdot\overrightarrow{R}_{i}}
   +\alpha m_{o}e^{i\overrightarrow{Q}_{o}\cdot\overrightarrow{R}_{i}}+
   \sigma\alpha m_{so}e^{i(\overrightarrow{Q}_{s}
   +\overrightarrow{Q}_{o})\cdot\overrightarrow{R}_{i}}].
\end{eqnarray}
For the $\alpha$ orbitals (1=zx and 2=yz) in iron pnictides, and for
the electrons in $\beta$ orbitals with $\beta=3
:x^{2}-y^{2},4:xy,5:3z^{2}-r^{2}$, we adopt,
\begin{eqnarray}
   <n_{i\beta\sigma}> &=& \frac{1}{2}[n_{\beta}+\sigma m_{\beta}e^{i
   \overrightarrow{Q}_{s}\cdot\overrightarrow{R}_{i}}]
\end{eqnarray}
where we define the particle numbers
$n_{12}=\frac{1}{N}\sum_{k\sigma, \alpha=1,2}
<C_{k\alpha\sigma}^{+}C_{k\alpha\sigma}>$,
$n_{\beta}=\frac{1}{N}\sum_{k\sigma}
<C_{k\beta\sigma}^{+}C_{k\beta\sigma}>$, and the SDW/ODW order
parameters $m_{s}=\frac{1}{N}\sum_{k\sigma,\alpha=1,2}\sigma
<C_{k\alpha\sigma}^{+}C_{k-Q_{s}\alpha\sigma}>$,
$m_{o}=\frac{1}{N}\sum_{k\sigma,\alpha=1,2}\alpha
<C_{k\alpha\sigma}^{+}C_{k-Q_{o}\alpha\sigma}>$,
$m_{so}=\frac{1}{N}\sum_{k\sigma,\alpha=1,2}\sigma\alpha
<C_{k\alpha\sigma}^{+}C_{k-Q_{s}-Q_{o}\alpha\sigma}>$,
$m_{\beta}=\frac{1}{N}\sum_{k\sigma}
\sigma<C_{k\beta\sigma}^{+}C_{k-Q_{s}\beta\sigma}>$, respectively.
In the momentum space the decoupled effective Hamiltonian can be
written as
\begin{eqnarray}
   H_{0} &=&
   \sum_{\substack{k,\alpha,\beta,\sigma}}[T^{\alpha\beta}(
   \overrightarrow{k}) C_{k\alpha\sigma}^{+}C_{k\beta\sigma}
   +T^{\beta\alpha}(\overrightarrow{k}) C_{k\beta\sigma}^{+}
   C_{k\alpha\sigma}-\mu C_{k\alpha\sigma}^{+}C_{k\alpha\sigma}]
\end{eqnarray}
and
\begin{eqnarray}
   \tilde{H_{I}} &=&
   \sum_{\substack{k,\sigma\\ \alpha=zx,yz}}[(A C_{k\alpha\sigma}^{+}C_{k\alpha\sigma}
   +\sigma B C_{k,\alpha\sigma}^{+}
   C_{k-Q_{s}\alpha\sigma}\\ \nonumber
   &&+\alpha C C_{k\alpha\sigma}^{+}C_{k-Q_{o}\alpha\sigma}+\sigma\alpha D
   C_{k\alpha\sigma}^{+}C_{k-Q_{s}-Q_{o}\alpha\sigma})]\\ \nonumber
   &&+\sum_{\substack{k\sigma\\ \beta=x^{2}-y^{2},xy,3z^{2}-r^{2}}}[
   (E_{\beta} C_{k\beta\sigma}^{+}C_{k\beta\sigma}
   +\sigma F_{\beta} C_{k\beta\sigma}^{+}
   C_{k-Q_{s}\beta\sigma})]
   +const.
\end{eqnarray}
where the coefficient
$A=\frac{1}{4}n_{12}U+\frac{1}{2}(\frac{1}{2}n_{12}+n_{3}+n_{4}+n_{5})(2U'-J_{H})$,
$B=-\frac{1}{4}m_{s}U-\frac{1}{2}(\frac{1}{2}m_{s}+m_{3}+m_{4}+m_{5})J_{H}$,
$C=\frac{1}{4}m_{o}[U-2U'+J_{H}]$,$D=-\frac{1}{4}m_{so}[U-J_{H}]$,
$E_{\beta}=\frac{1}{2}n_{\beta}U+\frac{1}{2}(n_{12}+n_{\bar{\beta}})(2U'-J_{H})$,
and
$F_{\beta}=-\frac{1}{2}m_{\beta}U-\frac{1}{2}(m_{s}+m_{\bar{\beta}})J_{H}$.
The constant term reads
\begin{eqnarray}
   const. &=&
   -\frac{N}{4}U[\frac{1}{2}(n^{2}_{12}+m^{2}_{o}-m^{2}_{s}-m^{2}_{so})+n^{2}_{3}+n^{2}_{4}+n^{2}_{5}
   -m^{2}_{3}-m^{2}_{4}-m^{2}_{5}] \\ \nonumber
   &&-NU^{'}[\frac{1}{4}(n^{2}_{12}-m^{2}_{o})+n_{12}(n_{3}+n_{4}+n_{5})
   +n_{3}n_{4}+n_{3}n_{5}+n_{4}n_{5}]\\ \nonumber
   &&+\frac{N}{2}J_{H}[\frac{1}{4}(n^{2}_{12}+m^{2}_{s}-m^{2}_{o}-m^{2}_{so})+n_{12}(n_{3}+n_{4}+n_{5})
   +n_{3}n_{4}+n_{3}n_{5}+n_{4}n_{5}\\ \nonumber
   &&+m_{s}(m_{3}+m_{4}+m_{5})+m_{3}m_{4}+m_{3}m_{5}+m_{4}m_{5}]
\end{eqnarray}

The e-l interaction describing the orthorhombic distortion is expressed as
\begin{eqnarray}
   H_{e-l}=
   \frac{\delta}{2}\sum_{\substack{k,\sigma}}(C_{k zx \sigma}^{+}C_{k-Q_{o} zx \sigma}
   -C_{k yz \sigma}^{+}C_{k-Q_{o} yz \sigma})+\frac{N\lambda}{2}\delta^{2}
\end{eqnarray}
 Minimizing the groundstate energy with
respect to the energy level splitting gives rise to the level splitting $\delta=
-\frac{1}{2N\lambda}\sum_{k\alpha\sigma}\alpha <C_{k\alpha\sigma}^{+}C_{k-Q_{o}\alpha\sigma}>$.
Thus the CF splitting $\delta$ can be obtained self-consistently
by $\delta=-\frac{1}{2}m_{o}/\lambda$.

\section{RESULTS AND DISCUSSIONS}

In this section, we first present our studies for the two orbital
model, and then for the more realistic five orbital model of the
quasi-two-dimensional systems, such as 1111 phase (LaOFeAs, etc.),
within the mean-field approximation, finally extend to the
three-dimensional systems in the 122 phase (BaFe$_{2}$As$_{2}$,
etc.). The magnetic and orbital properties closely related with a
tetragonal-orthorhombic structural phase transition in the iron
pnictides are explored. In addition, the effect of the inter-player
coupling for BaFe$_{2}$As$_{2}$ is also considered in the five
orbital model.

\subsection{Two-Orbital Model}
We firstly present our results in the case of the two-orbital model
\cite{PRB77-220503}. The possible spin and orbital configurations
with modulation wave-vectors
Q$_{s}$/Q$_{o}$=(0,0),(0,$\pi$),($\pi$,0) and ($\pi$,$\pi$) in both
the tetragonal structure and the orthorhombic phase are studied.
Note that in the tetragonal phase, the orthorhombic distortion is
absent, i.e. $H=H_{0}+\tilde{H_{I}}$; for comparison, while in the
orthorhombic phase the orthorhombic distortion is present, i.e.
$H=H_{0}+\tilde{H_{I}}+H_{e-l}$. The J$_{H}$-U phase diagrams  are
obtained both for the tetragonal phase and for the orthorhombic one,
as shown in Fig. 1 (a) and (b). In the tetragonal case, the SAFM
metallic phase is a stable ground state with small magnetic moment
in the intermediate Coulomb interaction parameter region, and all
the orbital configurations are degenerate in the SAFM phase, as seen
in Fig. 1 (a). So the ground state is nearly a para-orbital or
orbital liquid phase. However in the orthorhombic phase, the e-l
coupling breaks the orbital degeneracy of the xz-orbit and the
yz-orbit, leading to a ferro-orbital (FO) configuration. Due to the
itinerant character of the electron, the orbital polarization in the
ODW state is small. Since the pure electronic interaction is
insufficient to contribute the FO-SAFM phase, the Jahn-Teller-type
orthorhombic distortion removing the xz- and yz-orbital degeneracy
results in a weak FO ordering phase.
While, in the large U parameter region, the OO is destroyed under
the strong Coulomb interaction, which favors electron occupation in
the xz and yz orbitals equally.
In addition, the FO phase absence of SDW appears in the weak Hund's
coupling region, which is the result of the competition between the
crystal field splitting and the Hund's copling.
Here, the elastic coefficient parameter $\lambda$ is fixed
to be 0.5/t (t is the hopping integral parameter in Ref. \cite{PRB77-220503})
for the two-orbital model throughout this paper.

\begin{figure}[htbp]\centering
\includegraphics[angle=0, width=1.0 \columnwidth]{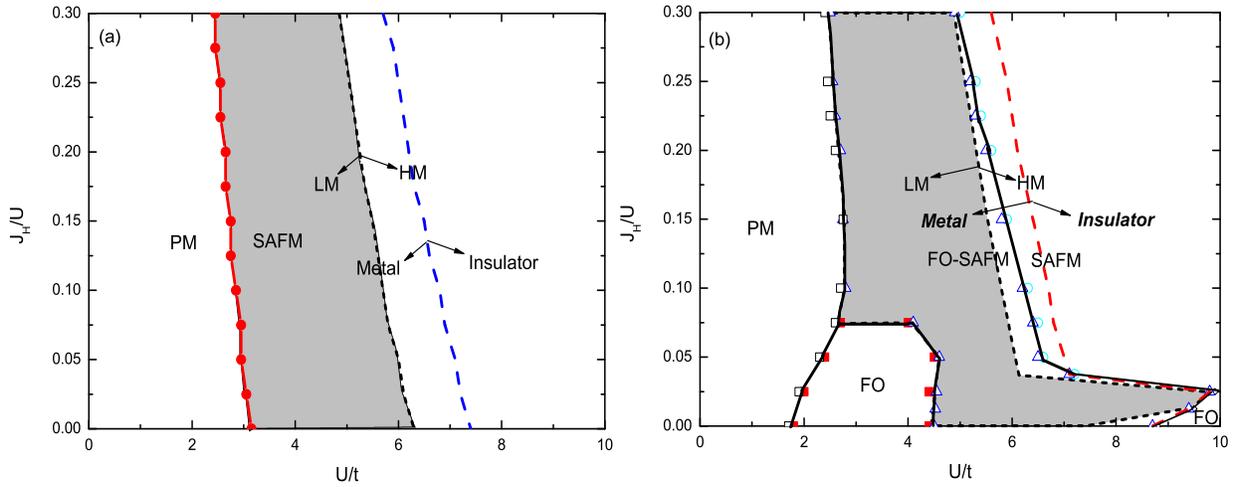}
\caption{J$_{H}$-U phase diagram of the two-orbital model in the
tetragonal (a) and orthorhombic (b) phases. PM and SAFM denotes
paramagnetic and striped antiferromagnetic ($Q_{s}$=$(\pi,0)$)
phase, respectively. FO represents the ferro-orbital order with
$Q_{o}$=$(0,0)$. The solid lines with different symbols display
different phase boundaries. The dash line and short dash line denote
the the metal-insulator border and the low magnetic (LM)
moment($\mu$ $<$ 1 $\mu_{B}$) and high magnetic (HM) moment ($\mu$
$>$ 1 $\mu_{B}$) phase border, respectively.
} \label{fig1}
\end{figure}

Minimizing the groundstate energy, we self-consistently obtain the
orthorhombic CF splitting, $\delta=-\frac{1}{2}m_{o}/\lambda$, for
the orthorhombic case. In Fig. 2, the dependence of the CF splitting
between the xz- and yz-orbit on the elastic coefficient $\lambda$ is
plotted with the Coulomb repulsion U=$4t$ for different Hund's
coupling J$_{H}$. For small coefficient $\lambda$, the CF splitting
$\delta$ becomes larger under the strong e-l interaction. However,
the large Hund's coupling obviously suppresses the CF splitting,
since the Hund's coupling tends to distribute the electrons in two
orbitals equal-weightly. This is also seen in Fig.1(b), the FO phase
only exists for small J$_{H}$.
\begin{figure}[htbp]\centering
\includegraphics[angle=0, width=0.5 \columnwidth]{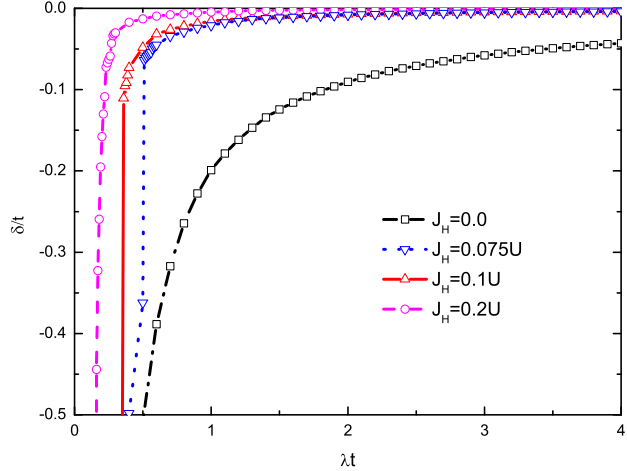}
\caption{Dependence of the crystal field splitting $\delta$ in the orthorhombic case
on the elastic coefficient $\lambda$ in two-orbital model, with the
parameters U=4t for J$_{H}$=0, 0.075U ,0.1U and 0.2U.}
\label{fig2}
\end{figure}

It is known that the SDW, ODW and combined spin-orbital density wave
order parameters, $m_{s}$, $m_{o}$ and $m_{so}$, crucially depends
on the Coulomb interaction. The comparison between the tetragonal
and the orthorhombic phases are displayed in Fig. 3. Although the
ordering parameters are rather small, an obvious orbital
polarization occurs in the orthorhombic case in the intermediate
Coulomb interaction range. As a comparison, it is absent in the
tetragonal case. A rather small SAFM magnetic moment is also
obtained in the intermediate coupling region, as seen in the shadow
region of Fig. 3. Notice that with the increase of U, the crossover
of the orbital polarizations from the xz-type symmetry to the
yz-type one occurs  at U$_{c}$ $\sim$ 4.5t.

\begin{figure}[htbp]\centering
\includegraphics[angle=0, width=0.6 \columnwidth]{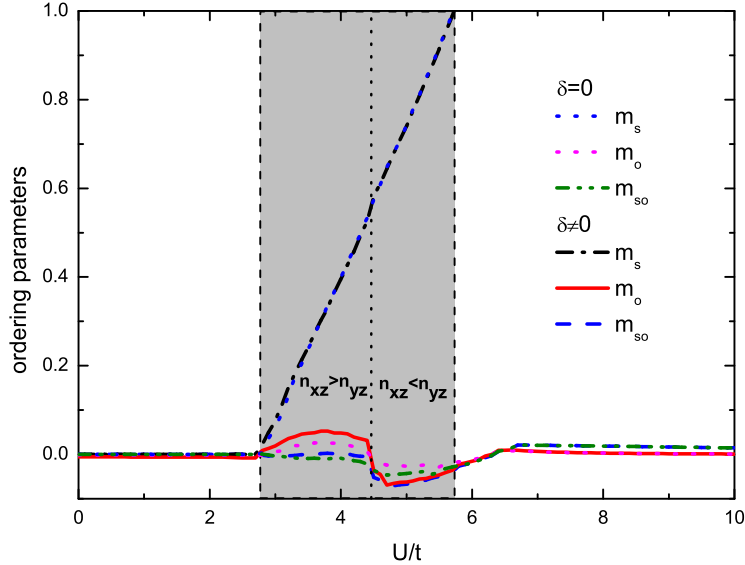}
\caption{Dependence of the spin, orbital and spin-orbital ordering
parameters on Coulomb interaction in the tetragonal ($\delta$=0) and
orthorhombic ($\delta$$\neq$0) phases, with J$_{H}$=0.1U. The
vertical dashed dotted line denotes the boundary of different
orbital polarization. The shadow region represents the small
magnetic moment case ($<$ 1 $\mu_{B}$).} \label{fig3}
\end{figure}

To compare partial density of state (PDOS) in both tetragonal and
orthorhombic cases are plotted in Fig. 4. It is shown that there is
no orbital polarization in the non-interacting case at U=0. With the
increase of U to a region of U$_{1}$$\sim$2.8t$<$U$<$U$_{c}$, the
PDOS of the xz-orbital component is larger than that of the
yz-orbital component near Fermi surface in orthorhombic case,
indicating an obvious xz-orbital polarization.
However, further increasing U to the region of
U$_{c}$$<$U$<$U$_{2}$$\sim$5.8t, it changes to a weak yz-orbital
polarization, in agreement with Fig. 3.
In the orthorhombic phase, there exists larger orbital polarization
than that in the tetragonal phase.
Obviously, although the electronic interaction plays a key role in
the orbital polarization, the e-l interaction finally stabilizes the
ground state of the LaOFeAs to the FO-type SAFM phase.
\begin{figure}[htbp]\centering
\includegraphics[angle=0, width=1.0 \columnwidth]{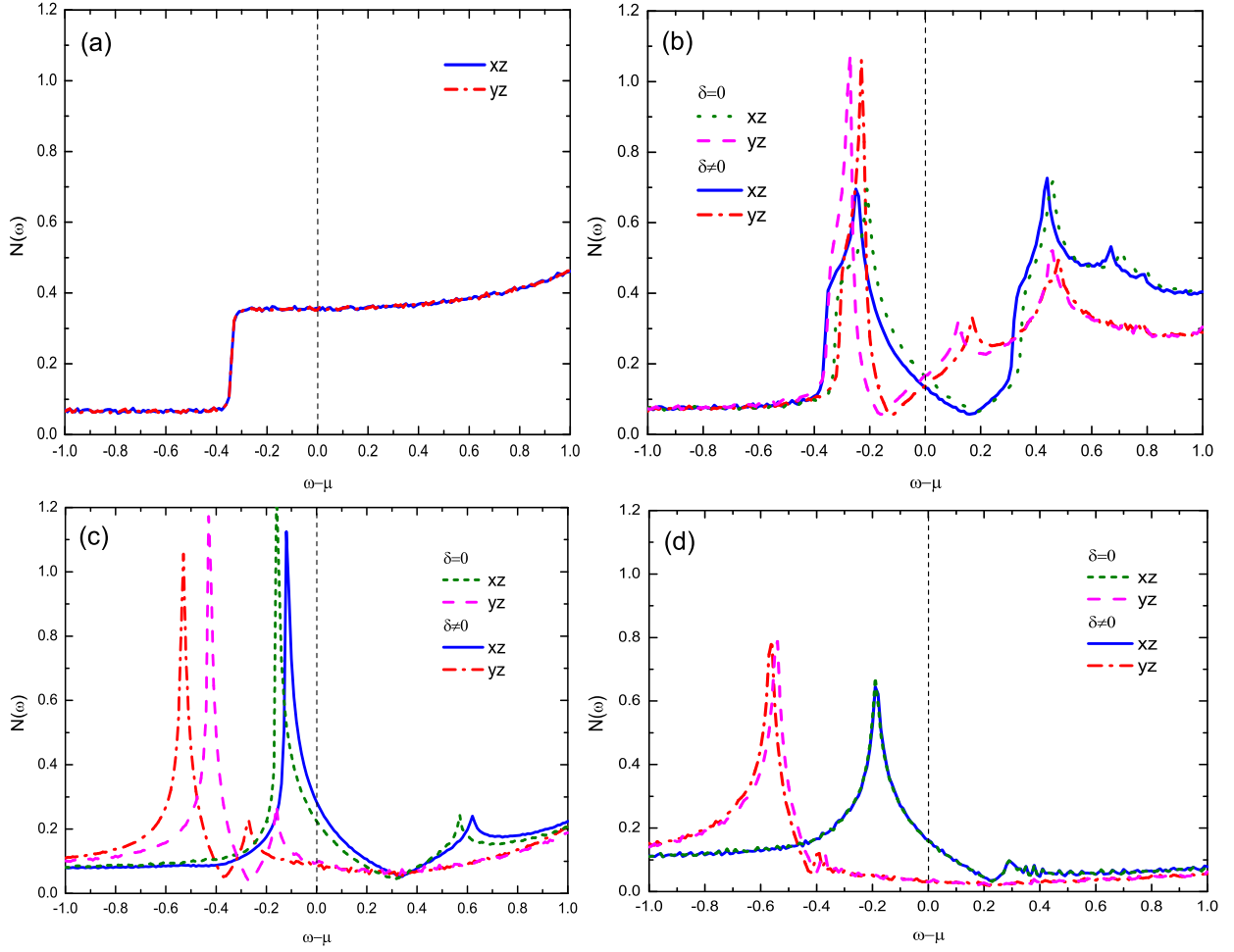}
\caption{Partial density of states in the tetragonal ($\delta$=0)
and orthorhombic ($\delta$$\neq$0) structures for different Coulomb
parameters  U=0 (a), 4t (b), 5t (c), and 6t (d) with $J_{H}$=0.1U.}
\label{fig4}
\end{figure}

To get a further insight into the SDW and ODW states, we display the
evolution of the Fermi surface on the Coulomb interaction in the
folded Brillouin zone of SAFM phase in Fig. 5. The Fermi surface
nesting is obviously observed in the non-interacting paramagnetic
case, including two hole pockets at the center and two electron
pockets at the corners of the Brillouin zone.
Nevertheless, with the increasing of the electronic
interaction, the nesting becomes weak and the ordered SAFM state is stabilized.

In addition, both the hole doping and the electron doping behaviors
are investigated in this two-orbital model, the results similar to
Ref.\cite{JPSJ78-083704} are obtained. Comparing with the
experiments, we find that in the hole doped case, the magnetic
moment decreases with doping, and is concordant with the
experimental results \cite{EPL83-27006}. While the AFM moment
increases with the increase of electron doping, which is completely
in contradiction to the experimental observations
\cite{EPL83-27006}.
On the other hand, the Fermi surface of the ordered state and the
xz-orbital polarization character of the Fermi surface in the
two-orbital model are not in agreement with the recent ARPES
experiments \cite{PRL104-057002}.
To resolve these discrepancies, we have to extend the present
effective two-orbital model to the realistic five-orbital model.
%
\begin{figure}[htbp]\centering
\includegraphics[angle=0, width=0.6 \columnwidth]{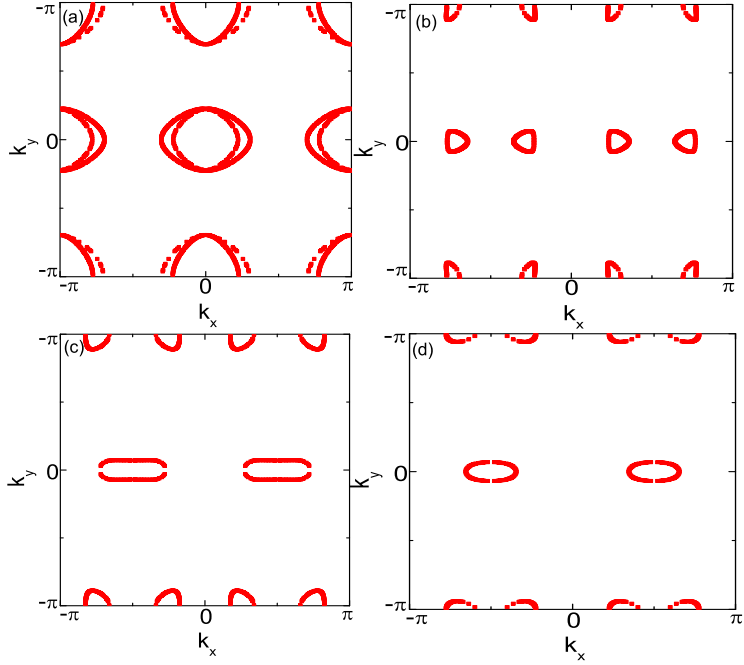}
\caption{Evolution of Fermi surface of the two-orbital models in the
folded Brillouin zone (a) U=0, (b) 4t, (c) 5t, and (d) 6t with $J_{H}$=
0.1U for the orthorhombic phase.} \label{fig5}
\end{figure}


\subsection{Five Orbital Model}

It is now generally believed that the more orbitals other than two
orbitals are involved in the low-energy physics in iron pnictides.
The five orbital of 3d electrons of Fe$^{2+}$ in the tetrahedral CF
should be considered to address the electronic, magnetic and orbital
properties of the iron pnictides. In this subsection, we explore the
role of the e-l coupling within the five-orbital model
\cite{NJP11-025016}. Notice that, unless otherwise specified, the
elastic coefficient parameter $\lambda$ is fixed to be 1.0 eV$^{-1}$
for the five-orbital model throughout this paper. Phase diagram of
the five-orbital model for the tetragonal (a), and orthorhombic (b)
phases are shown in Fig. 6. It is clearly found that the FO phase is
in a stable ground state in the presence of the orthorhombic
distortion. The intermediate Coulomb interaction with a weak Hund's
coupling favors the SAFM phase with low magnetic moments, similar to
the two-orbital model. However, the five-orbital results are
obviously different from that of the two-orbital model in Fig.1(a)
and Fig.1(b). With the increase of the electronic Coulomb
interaction, the N$\acute{e}$el AFM phase with $Q_{s}$=$(\pi,\pi)$
is in the ground state.
Moreover, in strong Coulomb repulsion and large Hund's coupling
region, the ground state of the system is the FO-SAFM phase with
large magnetic moment ($\mu$$>$1 $\mu_{B}$) due to the Hund's
coupling. This result is obviously different from the one without FO
in the simple two orbital case, which implies the five orbital case
involves a complicated multi-orbital correlated effect.
Further increasing the Coulomb correlation leads the system transit to insulator.
\begin{figure}[htbp]\centering
\includegraphics[angle=0, width=1.0 \columnwidth]{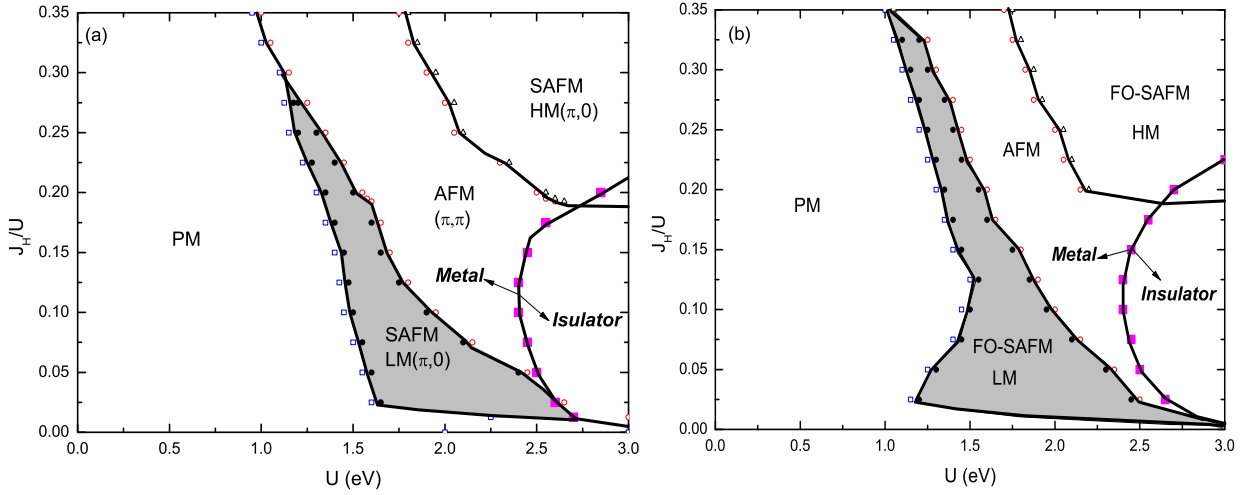}
\caption{J$_{H}$-U phase diagram of the five-orbital model
for the tetragonal (a) and orthorhombic (b) phases, respectively.
PM, SAFM and AFM denotes paramagnetic, striped antiferromagnetic ($Q_{s}$=$(\pi,0)$)
and N$\acute{e}$el antiferromagnetic ($Q_{s}$=$(\pi,\pi)$) phases, respectively.
FO represents the ferro-orbital order with $Q_{o}$=$(0,0)$.
LM and HM denote the low magnetization ($\mu$$<$1 $\mu_{B}$) and
high magnetization ($\mu$$>$1 $\mu_{B}$), respectively.} \label{fig6}
\end{figure}

The dependence of the magnetization of each orbital, and the orbital
occupancy, on the Coulomb interaction in both the tetragonal and the
orthorhombic structure are obtained, the orthorhombic case is shown
in Fig. 7. One finds that the xz-orbital polarization is larger than
that in the tetragonal one in a very narrow intermediate Coulomb
repulsion region from U=1.2 to 1.45 eV for $J_{H}$=0.25U. In this
region, the low magnetization, 0$<$$\mu$$<$1 $\mu_{B}$, with bad
metallic state is also obtained. The electron occupancy in the
xz-orbit is always more than that in the yz-orbit as the Coulomb
interaction increases. No orbital occupancy crossover from the
xz-type to the yz-type symmetry is observed, which is different from
the two-orbital results above. This shows that the ferro- xz-orbital
polarization arises from the electronic interaction of the five 3d
orbitals, while the long range ordering, i.e. OO is contributed from
the e-l interaction. Our results are consistent with not only the
crystal structure but also the ARPES experimental results
\cite{PRL104-057002}, where the xz-orbital polarization is observed
near the $\Gamma$ point.
\begin{figure}[htbp]\centering
\includegraphics[angle=0, width=1.0 \columnwidth]{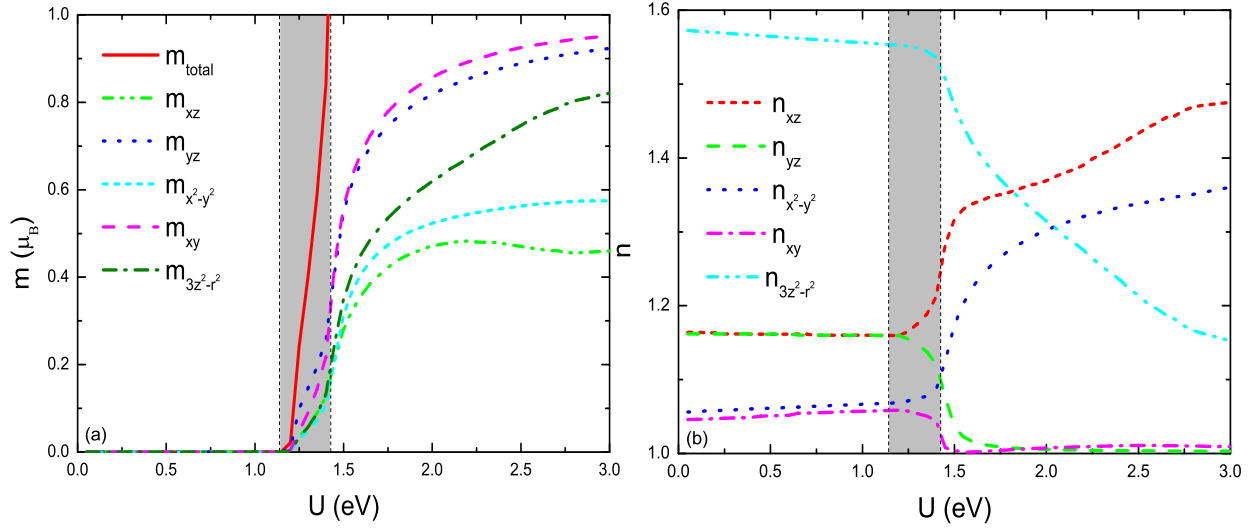}
\caption{Dependence of total magnetic moment, orbital magnetization
(a), and orbital occupancy (b) on Coulomb interaction in the
orthorhombic structure for J$_{H}$=0.25U. The shadow region denotes
the small magnetic moment physical parameters region with $\mu$$<$1
$\mu_{B}$.} \label{fig7}
\end{figure}

In the present five-orbital model, the orthorhombic CF splitting of
the xz- and yz-orbit is also determined self-consistently. The
dependence of the CF splittings on the e-l coupling, shown in Fig.
8, is similar to that in two-orbital situation. On the other hand,
in contrast to the two-orbital case, where the Hund's coupling
suppresses the CF splitting, the influence of the Hund's coupling on
the CF splitting is complicated in the five-orbital model because of
the multi-orbital effects. Either the strong or the weak Hund's
coupling J$_{H}$ favors the large orthorhombic CF splitting. Hence,
it indicates that the orthorhombic distortion is associated with the
multi-orbital effect. Thus, due to the xz-orbital polarization with
the ODW of wave vector $(0,0)$, it suffers an orthorhombic
distortion, leading to that the lattice parameter $a$ is larger than
$b$, as experimentally observed.
\begin{figure}[htbp]\centering
\includegraphics[angle=0, width=0.6 \columnwidth]{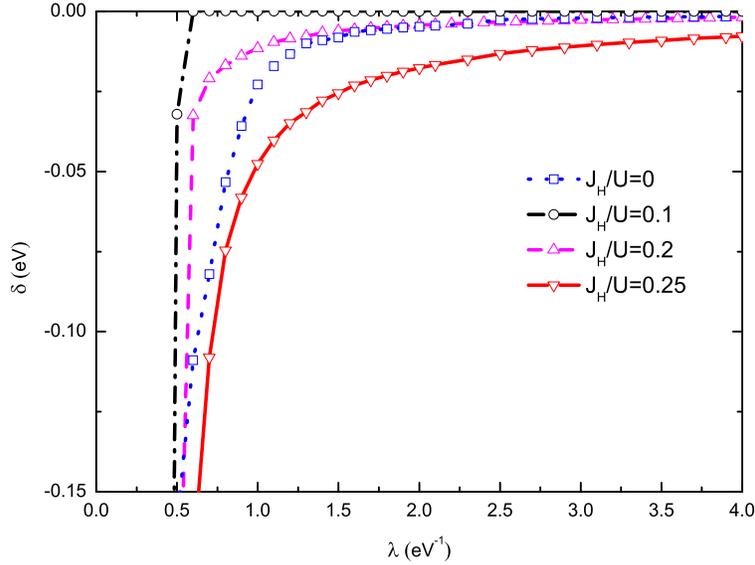}
\caption{Dependence of the crystal filed splitting of $xz$ and $yz$ orbitals
on the elastic coefficient with U=1.4 eV for different
Hund's rule coupling. Other parameters are the same with Fig.7} \label{fig8}
\end{figure}

In Fig. 9, the PDOS are plotted for various magnetic configurations:
non-magnetic, FO-SAFM(LM), N$\acute{e}$el AFM without OO, and
FO-SAFM(HM), respectively. In the non-interacting (a) and FO-SAFM
(b) cases, three t$_{2g}$ orbitals, xz, yz and xy components, mainly
contribute to the FS, which agrees with the LDA calculations
\cite{PRL100-226402} and the ARPES experiments \cite{PRB83-054510}.
In non-magnetic case, there is no orbital polarization of xz and yz
orbitals in FS. While in the small magnetic moment phase, the PDOS
of the xz-component is larger than that of the yz-component,
indicating the orbital polarization is mainly xz-component, as seen
in Fig. 9(b). In the N$\acute{e}$el AFM phase with
Q$_{o}$=$(\pi,\pi)$, the $3z^{2}-r^{2}$ orbital mainly contribute to
the FS. On the other hand, in the FO-SAFM(HM) phase, there is also
an obvious xz-orbital polarization near the FS. Notice that due to
the large magnetic moment and strong Coulomb interaction, the
electrons are localized electrons rather than itinerant electrons in
the SAFM states.
\begin{figure}[htbp]\centering
\includegraphics[angle=0, width=0.8 \columnwidth]{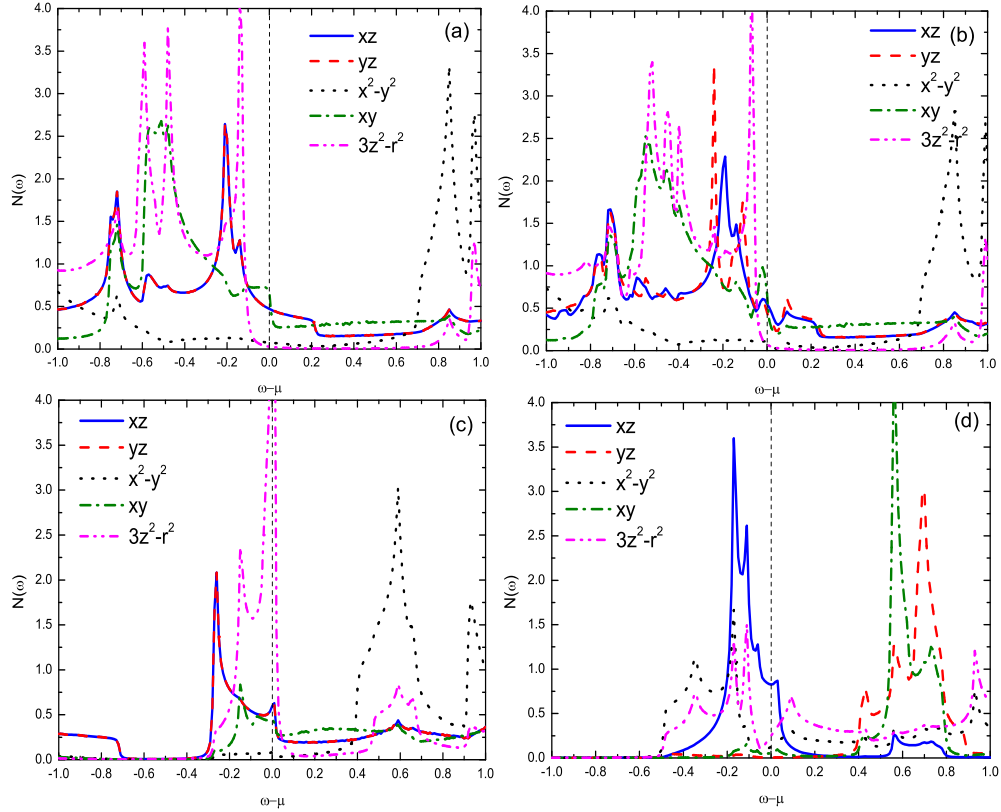}
\caption{Partial density of states in the paramagnetic (a), FO-SAFM(LM) (b),
N$\acute{e}$el AFM (c), and FO-SAFM (HM) (d) phases in the orthorhombic phase,
with parameters U=0, 1.25, 1.75 and 2.25 $eV$,
respectively. The Hund's coupling J$_{H}$=0.25U is adopted.} \label{fig9}
\end{figure}

With the increase of the Coulomb interaction, the evolution of the
FS is displayed in Fig. 10. From the PM state to the FO-SAFM state,
the FS nestings are destroyed with the increasing electronic
correlations. However, in the {\it N$\acute{e}$el} AFM states, the
FS nesting remains, which leads to the $(\pi,\pi)$ AFM states. With
the further increasing strong Coulomb interaction, the FS nestings
become weak.
Especially, the anisotropy of the FS appears in the SDW/ODW state at
the Coulomb interaction U$\sim$1.25 eV, with a larger hole pockets
along $\Gamma$$-$$Y$(0,$\pi$) in the folded SDW/ODW Brillouin zone
corresponding to the ferro-magnetic direction and a smaller one
along $\Gamma$$-$$X$($\pi$,0) corresponding to the
anti-ferromagnetic direction in the SAFM state.
The anisotropic character of the FS resembles the results of the
recent ARPES experiments \cite{arXiv1011.0050}.
The anisotropic FS is obviously different from the FS observed in
other ARPES experiments \cite{arXiv1011.1112} and the de Haas-van
Alphen experiment \cite{arXiv1009.1408}.
Furthermore, with the increase of U to 1.4 eV, the anisotropy
becomes weak, with the FS along $\Gamma$$-$$Y$ direction
disappearing. Our results reveal that the anisotropic character of
the FS and other properties are mainly due to the splitting of xz
and yz orbitals in the orthorhombic phase. Hence the anisotropy is
obviously manifested by the e-l coupling.
%
%

\begin{figure}[htbp]\centering
\includegraphics[angle=0, width=1.0 \columnwidth]{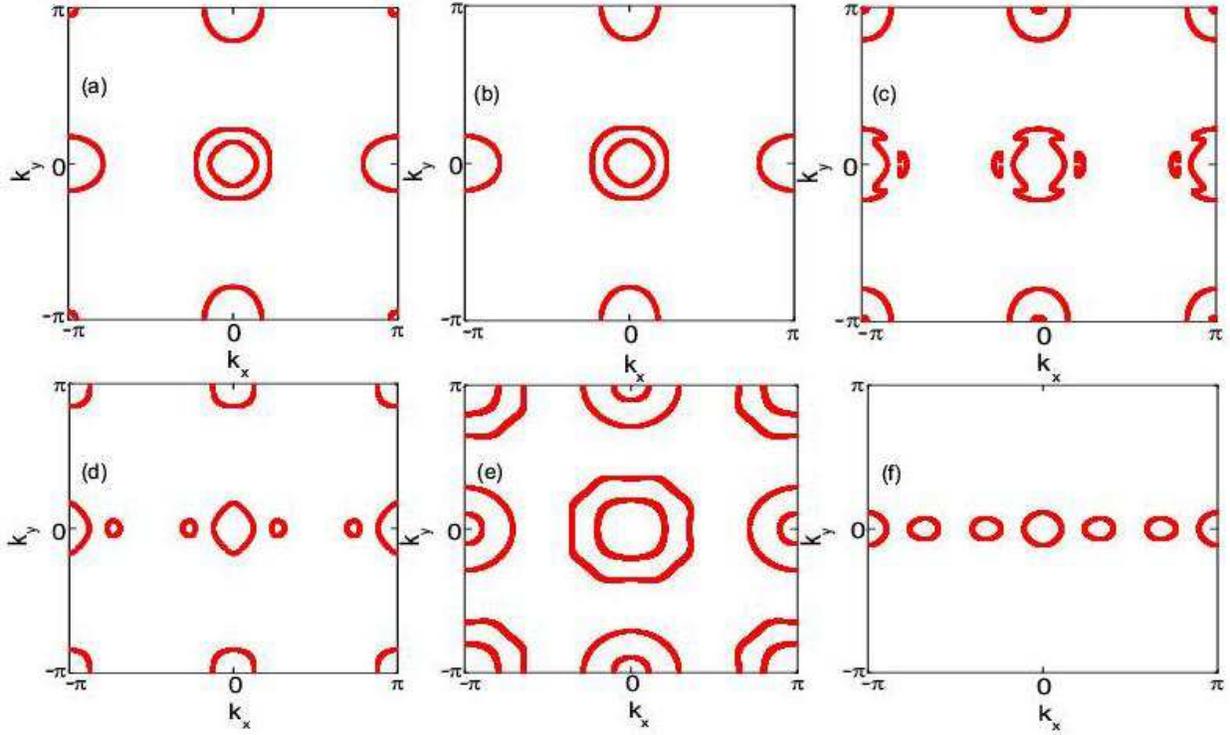}
\caption{Evolution of Fermi surface in the nonmagnetic/paramagnetic
(a) and (b), FO-SAFM of LM (c) and (d), N$\acute{e}$el AFM (e), and
FO-SAFM of LM (f) phases in the orthorhombic phase, with the Coulomb
correlation parameters are U=0 (a), 1.0 (b), 1.25 (c), 1.4 (d), 1.75
(e), and 2.25 (f) eV. J$_{H}$=0.25U is adopted.} \label{fig10}
\end{figure}

To compare with the two-orbital model, we also investigate the
doping case in the five-orbital model. We find that either hole or
electron doping results in the decrease of the magnetic moment,
which is well consistent with the experimental observations
\cite{EPL83-27006}. It is also implied that the five-orbital model
is a more accurate realistic model which well describes the
low-energy physics in the iron pnictides.

\section{SUMMARY}


In this work, we also study the realistic five-orbital model
\cite{PRB81-214503} for BaFe$_{2}$As$_{2}$ of more 3D systems, with
the SDW vector $Q_{s}$=($\pi$,0,$\pi$), and ODW vector
$Q_{o}$=(0,0,0) observed experimentally. To investigate the
relationship between the structural and magnetic phase transition,
the dependence of the magnetization and the ODW ordering parameters
on temperature in the tetragonal and orthorhombic structures is
observed. Note that the evolution of sublattice magnetization on
temperature reflects the variation of the spin alignments and the
long range order,. and the sublattice magnetization vanishes at the
critical temperature, characterized by T$_{SDW}$; while the
evolution of the OO parameters associated with the orthorhombic CF
splitting reflects the structural phase transition, which is
characterized by T$_{s}$. In absence of the orthorhombic distortion,
the pure electronic interaction leads to $T_{s}<T_{SDW}$, contrast
to the experimental results. However, in the presence of the
orthorhombic distortion, $T_{s}\gtrsim T_{SDW}$.
Therefore, in addition to the e-e interaction, addressing the e-l
interaction associated with the orthorhombic distortion clearly is
crucial to explain the magnetic and structural phase transitions
observed experimentally. Our results support a scenario that the
magnetic phase transition is driven by the ODW which is mainly
induced by the orthorhombic lattice distortion in the presence of
the intermediate electronic correlation.


In summary, we have presented the spin and orbital polarizations, as
well as SDW and ODW ordering, in both the tetragonal and
orthorhombic phase, and uncovered the relation between the
structural and magnetic phase transition. We have shown that the
orthorhombic lattice distortion is an essential factor to stabilize
the SAFM ground state through the formation of ODW ordering.

\acknowledgements

This work was supported by the NSFC of China No. 11047154, 11074257,
10947125, and the Knowledge Innovation Program of the Chinese
Academy of Sciences, and Natural Science Foundation of Anhui
Province No. 11040606Q56. Numerical calculations were performed at
the Center for Computational Science of CASHIPS.


\end{document}